\documentclass[useAMS,usenatbib]{mnras}
\usepackage{newtxtext,newtxmath}
\usepackage[T1]{fontenc}

\usepackage{graphicx}
\usepackage{times}

\begin{document}
\label{firstpage}
\pagerange{\pageref{firstpage}--\pageref{lastpage}} \pubyear{2024}

\newcommand{\msun}{$M_\odot$}
\newcommand{\masyr}{\,mas\,yr$^{-1}$\,}
\newcommand{\kms}{\,km\,s$^{-1}$\,}

\newcommand{\bfit}[1]{\mbox{\boldmath $#1$}}
\newcommand{\bfsf}[1]{\mbox{\boldmath \sf #1}}

\title[Orbits of triple systems]{Orbits and masses in two triple systems}

\author[Nazor \& Tokovinin]{Dinko Nazor$^{1}$\thanks{E-mail:dinko.nazor@gmail.com} 
and Andrei Tokovinin$^2$\thanks{E-mail:andrei.tokovinin@noirlab.edu} \\
$^1$  Pixel Ribs d.o.o. - Vla\v{s}ka 79, 10000 Zagreb, Croatia \\
$^2$ Cerro Tololo Inter-American Observatory -- NSFs NOIRlab, Casilla 603, La Serena, Chile\\
}

\date{-}

\maketitle

\begin{abstract}
In an effort to determine  accurate orbital and physical properties of
a  large  number of  bright  stars,  a  method  was developed  to  fit
simultaneously  stellar  parameters (masses,  luminosities,  effective
temperatures), distance, and orbits to  the available data on multiple
systems, namely  the combined and differential  photometry, positional
measurements, radial velocities (RVs), accelerations, etc.  The method
is applied to a peculiar  resolved triple system HIP~86286. The masses
of its components  estimated  using observations and  standard
  relations are  1.3, 0.9,  and 0.9  \msun; the main  star is  a G8IV
subgiant,  while its  two  companions are  main-sequence dwarfs.   The
inner and  outer orbital periods  are 35 and 287  years, respectively,
and the orbits are nearly coplanar.  The second system, HIP 117258, is
an  accelerating star  with a  resolved companion;  its 35.7-yr  orbit
based on relative astrometry and precise RVs yields the secondary mass
of  0.95 \msun,  much larger  than inferred  from the  photometry. The
apparent  paradox is  explained by  assuming that  the secondary  is a
close pair of M-type dwarfs with yet unknown period.
\end{abstract}


%--
\section{Introduction}
\label{sec:intro}

Astronomy  is based  on observations,  and this  general statement  is
fully relevant  for the study  of stellar multiple systems.   Only the
brightest stars  were accessible to visual  examination or photographic
spectroscopy  in the  19th  century.  As  the astronomical  technology
progressed, larger samples of fainter objects became scrutinized.  The
Tycho mission on  board the Hipparcos satellite examined  all sky down
to  the  12th  magnitude.   Its  combination  with  prior  ground-based
astrometry resulted in  the Tycho-2 catalogue of 2.5  million stars with
accurate positions and proper motions (PMs) \citep{Tycho2}.  Nowadays,
the Gaia project pushes the limits much further by monitoring billions
of stars  down to  the 21st magnitude  \citep{DR3}.  For  the brighter
part of the Gaia targets,  astrometry is complemented by spectroscopy.
The number  of astrometric and  spectroscopic orbits published  in the
Gaia  catalogue of  non-single stars  \citep{NSS} already  surpasses the
totality of previously known  ground-based orbits, although the latter
cover a much wider range of periods.

Despite this tremendous progress, the brightest stars still possess the
most detailed and complete  observations. Furthermore, many systems of
two  or  more  stars  have   problematic  data  in  Gaia  for  various
reasons. In an effort  to  assemble the most up-to-date and  accurate data on
the  stars  of  the  Tycho  catalogue,  numerous  cases  of  missing  or
unsatisfactory   data  were   identified.    This  prompted   detailed
examination  of  observed motions  in  selected  multiple systems  and
determination or update of their orbits.  The purpose of this paper is
to share several results of this work undertaken by the first author
(DN),  a  programmer and  an  amateur  astronomer.   The role  of  the
co-author  (AT) is  to verify  the  results and  to present  them in  a
concise   way,   contributing   professional   experience   to   this
collaboration.

In  the epoch  of large  surveys, individual  examination of  multiple
systems  might appear  outdated. Yet,  most (if  not all)  surveys are
designed for single  stars; the presence of one  or several additional
components affects the  single-star pipelines and leads  to biases and
errors,  sometimes significant.   For example,  2/3 binary  stars with
known visual  orbits still  lack Gaia  parallaxes \citep{Chulkov2022}.
With the planned 11 yr duration  of the full Gaia mission, orbits with
longer  periods  have  to   rely  on  the  complementary  ground-based
observations.

The methods of orbit calculation  and assessment of stellar parameters
are   briefly   outlined   in  Section~\ref{sec:methods}.    Then   in
Section~\ref{sec:GJ} one interesting resolved triple system is studied
in  detail  to   illustrate  this  approach.  Section~\ref{sec:117258}
reveals another  unusual case,  namely a hidden  triple where  the dim
secondary is  in fact  a close  pair of low-mass  stars. The  paper is
summarized in Section~\ref{sec:sum}.

%-------------------------------------
\section{Data and Methods}
\label{sec:methods}

As a part of a greater  effort to validate and improve existing orbits
and compute  new ones,  DN has  written a  framework of  classes and
functions. The  language of  choice is C/C++  for its  well-known high
performance. The main  goal is to come up with  a unified, robust, and
dynamic   model   of  our   Galactic   neighbourhood.   There  is   no
``ready-to-use'' application where the input is dropped-in and results
are   spat-out.    Each  case   rather   needs   to  be   individually
programmed. This  may seem like  a drawback,  but the reality  is that
many  systems  require  a  particular intervention  that  wouldn't  be
possible   with    a   ready-to-use   software   and    without   some
fine-tuning. So, when  such situation is encountered,  usually a small
adaptation can make a big difference.

The general types of models that the framework supports are 2 to 4-star
systems  in  A-B,  Aab-B,  A-Bab,  and  Aab-Bab  configurations.   The
accepted  input  data  are   relative  positions,  RVs,
accelerations,  light-source  ``wobbling'',  light  curves,  times  of
minima for eclipsing  binaries, and observed magnitudes in  the $U, B,
V, R,  I, J, H, K,  G, G_{\rm BP}, G_{\rm  RP}, u, g, r,  i, z$ bands.
The  model represents  orbits, astrometry,  and stellar  parameters of
each  star  in over  60-dimensional  parameter  space.  It  is  highly
optimized for performance,  so for example, for systems  with more than
two components,  the light-time effect  can be enabled and  taken into
account with negligible additional computational cost.  Quadratic limb
darkening coefficients  for computing  magnitudes during  eclipses are
implemented  using  the tables  from  \citet{Claret2018}  to get  the
coefficients for a particular star  from its effective temperature and
gravity. The  model is  also very versatile.   For example,  it allows
different system velocities to be  assigned to multiple sets of RVs.
 Rectilinear solution is the  last resort for wide binaries
for which  the observed relative motion  has a short time  coverage or
has not exhibited sufficient changes.

For computing  stellar parameters,  DN has implemented  various models
for  different types  of  stars. The  tables from  \citet{Pecaut2013},
updated                                              online,\footnote{
  https://www.pas.rochester.edu/\~{}~emamajek/EEM\_dwarf\_UBVIJHK\_colors\_Teff.txt}
have proven to be  the best choice; for red stars  with $T_{\rm eff} <
5500$K, the  2019 version of  the online tables gives  slightly better
results than  the 2021 version.  The  BT-Settl \citep{Baraffe2015} and
PARSEC \citep{PARSEC} evolutionary tracks  are used for evolved stars.
 Please,  note that  stellar masses  and other  parameters derived
  here  are model-dependent.   Our goal  is to  find the  best set  of
  self-consistent parameters that fits the available data, rather than
  to test stellar evolutionary models. 

As usual, the solvers try to  find the global minimum of $\chi^2$. The
core solver  is based  on the  Gauss-Newton method  with Moore-Penrose
inverse, with a custom line-search algorithm. It is a numerical method
which finds  the local $\chi^2$  minimum,  given an  initial guess, down  to the
64-bit precision of floating-point numbers.

 The initial guess can  be either found by a combination
of MCMC \citep{MCMC} and simulated annealing,  or it can simply be (in
most  cases) a  known (good)  solution.  In  the former  case, several
different attempts  are constructed  and manually inspected  to verify
similarity  of  the convergences.   The  parameters  can be  fixed  or
ignored  during computation  (the  algorithm automatically  recognizes
parameters that are not affected by the input data and disables them),
and any additional specific constraints can be added easily via lambda
functions.  Automatic  computation of parameters can  also be included
in the process via lambda functions (e.g. when radii are fixed to fill
stars'  Roche lobes,  when the  mass-radius relationship  is for  some
reason strictly  determined, or  when limb darkening  coefficients are
computed from  the stellar parameters). The  Gauss-Newton solver might
fail if the inputs are ``too contradictory'' or if the solution is not
well  constrained or  defined. To  get around  this problem,  a simple
manual data  reduction or  assigning smaller  weights to  outliers are
helpful. The solver is rather fast; it converges within a few 10th of
a second (less than 100 iterations)  even with hundreds of positions or
thousands of light-curve measurements, for example.

%-------------------------------------
\section{The triple system HIP 86286 (GJ 683.2)}
\label{sec:GJ}

\begin{figure}
\includegraphics[width=7cm]{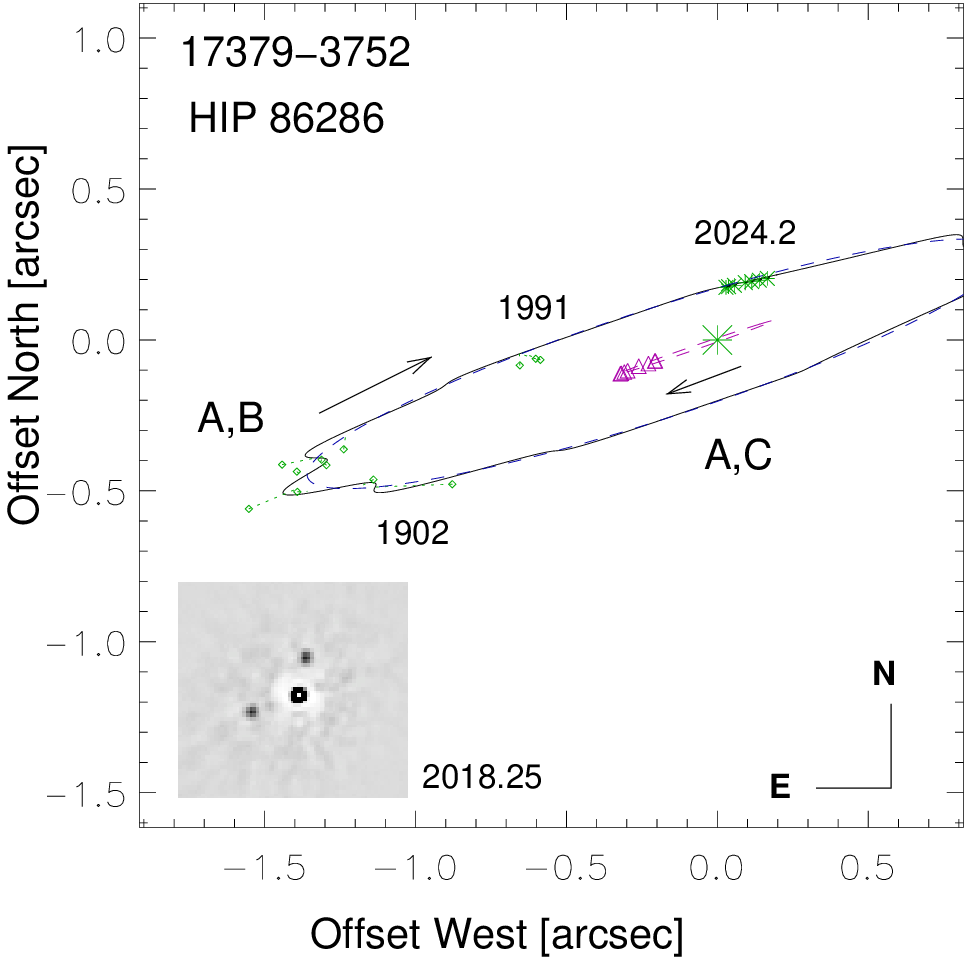}
\caption{Orbits  of  the  triple  system HIP  86286  (axis  scale  in
  arcseconds).  The  magenta triangles  and ellipse show  the measured
  positions and the orbit of the inner subsystem A,C.  The black solid
  line depicts  the outer orbit  with wobble caused by  the subsystem,
  while   the  blue   dashed   line  is   the   outer  orbit   without
  wobble.  Positions of  the outer  pair  are plotted  as small  green
  crosses and larger  green asterisks when they correspond  to AC,B and
  A,B, respectively.   The insert shows the  shift-and-add image taken
  at SOAR on 2018.25 and indicating true quadrants of the companions.
\label{fig:orb} 
}
\end{figure}

This bright  stellar system, known  as HIP~86286, HD~159704,  WDS J17379$-$3752,
and GJ~683.2,  was resolved in  1897 by R.~Innes  as a $1''$  pair; it
bears the  discoverer name  I~247.  The separation  between A  and its
faint companion B  gradually decreased during the  20th century.  This
neglected  pair was  observed  by  the speckle  camera  at  the 4.1  m
Southern  Astrophysical  Research  (SOAR)  telescope in  2017  at  the
request  by R.~Gould  and  found  to be  a  resolved  triple with  two
companions  B  and  C  of  comparable  brightness  and  at  comparable
separations \citep{SOAR2017}.  Such trapezium-type  configurations are
expected to  be dynamically  unstable.  The  perplexing ``triangular''
system was  re-visited at SOAR  with a yearly cadence,  revealing slow
motion of both companions in opposite directions (C is receding from A
and  B  is approaching).   AT  suspected  initially  that one  of  the
companions  could   be  unrelated   (the  sky   in  this   region  is
crowded). However, in such a case the fast PM of $(-5.2,
-131.5)$  \masyr \citep{HIP2}  would  have  changed the  configuration
rapidly, so both companions are actually physically bound to A.

In June 2021, DN stumbled  upon this interesting system while revising
information  on   Tycho  stars.   He  noted   the  strange  triangular
configuration  with comparable  separations  and  also concluded  that
companions  B and  C are  related to  A. Assuming  initially that  the
companion C was unrelated, he tried to compute the orbit of AB, but it
matched  the  measured  positions  very  poorly.   There  was  also  a
significant   discrepancy  between   the   Hipparcos   and  Gaia   DR2
PMs; the acceleration computed  from the provisional AB orbit
could not possibly explain the measured one of the primary. It clearly
suggested  that the  primary should  be accelerating  in the  opposite
direction! The  only explanation  was the  reflex motion  of A  in the
inner orbit AC.

A model for three-star system was set up and fed with the data. Due to
the shortness  of the observed  arcs and unavailability of  other data
(RVs),  the solution was still  purely for demonstrative
purposes. However, all relative positions could be closely matched and
the strange motion  of the primary star A was  reproduced well, within
the assumed errors. So, the object has revealed itself  beyond any
doubt as a triple system: a close binary AC with  a period of a few decades and
their outer companion B in a few hundred year orbit. Star A is an
evolved subgiant and both companions  are lighter and assumed to still
be  on the  main sequence.  

At this  point, the information was  shared with AT who  confirmed the
validity of the proposed triple-star  model.  A preliminary version of
these  orbits   is  listed  in   the  online  Multiple   Star  catalogue
\citep{MSC}.  The orbits  are determined here using the  latest (up to
2024.2) position measurements at SOAR.  The inner and outer orbits are
plotted in  Fig.~\ref{fig:orb}.  The  orbital elements,  position, and
motion of the centre of  mass are listed in Table~\ref{tab:orb}, while
stellar  parameters  are  given  in  Table~\ref{tab:par}.   They  were
determined by the joint fit  to positions and photometry, as explained
below.
 
\begin{table}
\center
\caption{Orbits and Astrometry of HIP 86286}
\label{tab:orb}
\medskip
\begin{tabular}{l   c  c} 
\hline
Element & A,C   & AC,B \\
\hline
$P$ (yr) & 34.95$\pm$0.05 & 287.0$\pm$1.4 \\
$T$ (JY) & 2041.907$\pm$0.022 & 2049.65$\pm$0.40 \\
$e$      & 0.270$\pm$0.022 & 0.192$\pm$0.005 \\ 
$a$ ($''$)& 0.2677  & 1.2204 \\
$\Omega$ (deg) & 109.05$\pm$0.20 & 107.86$\pm$0.22 \\
$\omega$ (deg) & 180.77$\pm$0.22 & 156.58$\pm$0.22 \\
$i$ (deg) & 88.54$\pm$0.22 & 98.69$\pm$0.07 \\
RA, Decl. (J2000) & \multicolumn{2}{c}{17:37:51.2284, $-$37:51:41.053} \\
$\mu^*_\alpha$, $\mu_\delta$ (\masyr) &  \multicolumn{2}{c}{$-$10.52,  $-$129.52} \\
Parallax (mas)    & \multicolumn{2}{c}{19.274$\pm$0.041} \\
\hline
\end{tabular}
\end{table}

\begin{table}
\center
\caption{Stellar Parameters and Photometry of HIP 86286}
\label{tab:par}
\medskip
\begin{tabular}{l   c  c c c} 
\hline
Parameter & A  & C  & B & A+B+C \\
\hline
%Parallax (mas)    & \multicolumn{4}{c}{19.274} \\
Mass (\msun)      & 1.28 & 0.91 & 0.89 & 3.08 \\
Radius ($R_\odot$) & 2.27 & 0.89 & 0.86 & -- \\
$T_{\rm eff}$ (K)   & 5421 & 5451 & 5370 & -- \\
$L$ ($L_\odot$)    & 4.01 & 0.624 & 0.552 & -- \\
$\log g$ (cm\,s$^{-2}$) & 3.83 & 4.50 & 4.52 & -- \\
Sp. Type          & G8.5IV & G8.3V & G8.8V & -- \\
$B$ (mag)         & 7.717 & 9.749 & 9.927 & 7.45 (7.45) \\
$V$ (mag)         & 6.959 & 9.007 & 9.161 & 6.69 (6.68) \\
$R_C$ (mag)       & 6.542 & 8.598 & 8.742 & 6.27 (6.28) \\
$I_C$ (mag)       & 6.151 & 8.212 & 8.742 & 5.88 (5.88) \\
$J$ (mag)         & 5.578 & 7.646 & 7.765 & 5.31 (5.36:)\\
$H$ (mag)         & 5.224 & 7.298 & 7.405 & 4.95 (5.12:)\\
$K_s$ (mag)       & 5.141 & 7.215 & 7.319 & 4.87 (4.87)\\
$G$ (mag)         & 6.752 & 8.815 & 8.960 & 6.48 (6.48)\\
$G_{\rm BP}$ (mag)  & 7.156 & 9.222 & 9.381 & 6.89 (6.90)\\
$G_{\rm RP}$ (mag) &  6.221 & 8.309 & 8.442 & 5.96 (5.95)\\
\hline
\end{tabular}
\end{table}

The input  data include the combined  photometry in bands from  $B$ to
$K_s$ (the  measurements are given in  brackets in the last  column of
Table~\ref{tab:par}).   The errors  of 0.1  mag are  assumed, and  the
uncertain 2MASS photometry  in the $J$ and $H$ bands  is not used. The
model reproduces  the combined  photometry very  well. The  model also
uses the differential photometry from  SOAR in the $I_C$ band: $\Delta
m_{AC} = 2.06 \pm 0.10$ mag, $\Delta  m_{AB} = 2.19 \pm 0.10$ mag. The
Gaia photometry  is taken from  DR2.  Zero interslellar  extinction is
adopted,  considering   the  small  distance,  and   confirmed  during
preliminary  computations.   Table~\ref{tab:pos}  lists  the  position
measurements,  residuals from  the  orbits, and  adopted errors.   The
position angles  $\theta$ are  corrected for  precession to  the epoch
2000.0,  the  separations $\rho$  are  given  in mas.   The  published
measurements at SOAR in 2017--2020  are corrected for minor systematic
errors as prescribed in \citep{Tok2022}.  Large errors are assigned to
the inaccurate  positions, effectively  canceling these  data.  Visual
measurements  of  AB-C  made  before  1991  were  retrieved  from  the
Washington Double Stars (WDS) database,  the measurements in 1991 come
from  the Hipparcos  and Tycho  missions, and  the remaining  data are
obtained at SOAR.  Note that the  companion's designations B and C are
swapped in the published SOAR data.

The errors of the orbital elements  are determined by running the MCMC
chains $10^4$ times and fitting  Gaussians to the distributions of the
resulting  elements,  which  are  nicely  bell-shaped.   However,  the
explored region of the parameter space was restricted by the condition
$\chi^2 < \chi^2_{\rm min} + 5$.  The MCMC did not return any $\chi^2$
smaller than the minimum $\chi^2_{\rm min}$ found by the Gauss-Newton
method.  The semimajor axis was not among the fitted parameters (it is
computed from  the masses, periods,  and parallax), so its  errors are
not given in Table~\ref{tab:orb}. AT  checked the results with the IDL
code   {\tt   orbit3}   that    fits   only   the   orbital   elements
\citep{ORB3}.  Similar elements,  but  with much  larger errors,  were
found.  So, the errors in Table~\ref{tab:orb} might be underestimated.

\begin{table}
\center
\caption{Positions and residuals of HIP 86286}
\label{tab:pos}
\medskip
\begin{tabular}{l   c cc ccc} 
\hline
Date     & $\theta_{2000}$ & (O-C)$_\theta$ & $\sigma_\theta$ &  $\rho$ & (O-C)$_\rho$ &  $\sigma_\rho$ \\ 
(JY)    & (deg) & (deg)   & (deg) & (mas) & (mas) & (mas) \\
\hline
\multicolumn{7}{c}{AC (inner pair)}         \\
2017.5345 &   108.6  &   $-$0.0  &   0.5  &      216.9  &  $-$0.3  &  2.0  \\ 
2017.5345 &   108.7  &     0.1  & 999.9  &      219.1  &    1.9  & 99.9   \\  
2018.2526 &   108.9  &     0.2  &   0.5  &      241.7  &    0.3  &  2.0   \\  
2019.3803 &   108.5  &   $-$0.3  &   0.5  &      274.5  &    0.3  &  2.0  \\   
2021.2452 &   108.9  &     0.0  &   0.5  &      314.6  &    0.3  &  2.0  \\   
2022.1954 &   109.0  &     0.1  &   0.5  &      326.4  &  $-$1.2  &  2.0  \\   
2023.3243 &   109.0  &     0.0  &   0.5  &      337.2  &    0.0  &  2.0  \\   
2024.2391 &   109.0  &   $-$0.0  &   0.5  &      340.4  &    0.5  & 2.0  \\ 
\multicolumn{7}{c}{AC-B (unresolved AC)} \\
1897.0009 &   119.3  &      6.1 &  999.9  &     1000.0 &  $-$234.1 & 500.0 \\   
1902.4308 &   112.1  &    $-$0.2 &    1.5  &     1230.0 &   $-$17.5 &  50.0  \\  
1914.6205 &   109.9  &    $-$0.4 &    1.5  &     1480.0 &      4.2 &  50.0  \\  
1920.5604 &   109.8  &      0.3 &    1.5  &     1650.0 &    132.1 & 200.0 \\   
1929.6102 &   107.4  &    $-$0.9 &    1.5  &     1460.0 &      6.0 &  50.0  \\  
1936.4801 &   106.9  &    $-$0.4 &    1.5  &     1370.0 &      9.0 &  50.0 \\   
1938.8400 &   107.8  &      0.8 &    1.5  &     1360.0 &      1.8 &  50.0 \\   
1943.2499 &   106.0  &    $-$0.4 &    1.5  &     1500.0 &    110.7 & 300.0  \\  
1959.6196 &   106.3  &      2.2 &    3.0  &     1290.0 &    $-$5.2 &  50.0  \\  
1991.1889 &    97.3  &      2.6 &    3.0  &      660.0 &   $-$30.4 &  50.0 \\   
1991.2500 &    96.0  &      3.5 &   10.0  &      605.0 &      7.9 & 100.0 \\   
1991.5700 &    96.5  &      4.3 &   10.0  &      590.0 &      0.8 &  100.0 \\
\multicolumn{7}{c}{A-B (resolved AC)} \\   
2017.5345 &   350.8  &  $-$0.3  &   0.5  &      176.3  &  $-$0.4 &   2.5   \\
2017.5345 &   351.8  &    0.7  & 999.9  &      176.1  &  $-$0.6 &  99.9   \\
2018.2526 &   348.1  &    0.3  &   0.5  &      180.2  &  $-$0.4 &   2.5   \\
2019.3803 &   342.7  &    0.1  &   0.5  &      189.4  &    0.4 &   2.5   \\
2021.2452 &   334.0  &    0.2  &   0.5  &      210.7  &    1.2 &   2.5   \\
2022.1954 &   329.5  &  $-$0.0  &   0.5  &      224.1  &    0.6 &   2.5   \\
2023.3243 &   324.9  &    0.1  &   0.5  &      241.8  &  $-$1.5 &   2.5   \\
2024.2391 &   320.9  &  $-$0.4  &   0.5  &      262.5  &    0.5 &   2.5   \\   
\hline
\end{tabular}
\end{table}

Stellar parameters of the main-sequence stars B and C are derived from
the online tables  of Mamajek (2019 version), while  the magnitudes of
the evolved star A are modeled  according to the 2021 version of these
tables.  The  masses $M$ of   dwarfs and  subgiants with $M  > 0.8$
  \msun are  computed from  effective temperature $T_{\rm  eff}$ and
luminosity $L$ by  the formula 19 from  \citet{Montalto2021}, based on
Table~10 of \citet{Moya2018}:
\begin{equation}
\log( M/M_\odot) = -0.119 + 2.14\,10^{-5} T_{\rm eff} + 0.1837 \log (L/L_\odot) .
\label{eq:mass}
\end{equation}
We  preferred to  use this  formula (rather  than tables)  because the
evolutionary tracks are not monotonous, precluding convergence in some
cases. For dwarfs less massive than  0.8 \msun, we use instead the 5th
order polynomial  from \citet{Mann2019} relating mass  to the absolute
magnitude $M_{Ks}$.   The code  fits simultaneously  orbits, effective
temperatures  $T_{\rm eff}$,  radii  $R$ of  all  components, and  the
parallax.   For  a  given  pair   of  $T_{\rm  eff},  R$  values,  the
luminosities and masses are computed;  the gravity $\log g$ is derived
from $M$ and $R$.  Using  the bolometric corrections and colours given
by  the  standard  relations,  the absolute  magnitudes  in  different
photometric  bands are  determined.  They  are translated  into actual
magnitudes by adding the distance modulus and extinction and fitted to
the available combined and differential photometry.  For single stars,
the parameters derived by this method are very close to those found in
the TESS Input  Catalogue \citep{TIC}.  The third Kepler  law is built
into the model,  so the semimajor axis of each  orbit is computed from
its period, mass sum, and parallax.

Our  model determines  the  parallax of  19.274 mas,  so  the star  is
definitely beyond  the 25  pc limit  of the  Gliese catalogue  of nearby
stars (it was  likely included there because of  the large photometric
parallax and a fast PM).  The space  astrometry of this triple system is
discordant:  Hipparcos  measured  a parallax  of  21.02$\pm$1.09  mas
\citep{HIP2},  Gaia  DR2 ---  16.33$\pm$0.46  mas,  and Gaia  DR3  ---
14.71$\pm$0.80  mas. The  large  RUWE  of 20.5  and  the  bias of  the
parallax  in  Gaia  are  likely   produced  by  displacements  of  the
photo-centre due  to the  unresolved companions C  and B.   This shift
depends  on the  scan direction,  which changes  regularly during  the
year, biasing eventually the parallax.  This source is challenging for
Gaia, and even the final data release might fail to give good results,
unless a triple-star model is fitted explicitly.

The PM of  the system derived from its positions  in the Hipparcos and
Gaia  DR3   catalogues  is   $(\mu^*_\alpha,  \mu_\delta)   =  (-10.517,
-129.521)$ \masyr  \citep{Brandt2021}. 
We  compared  the  PMs  measured   by  Hipparcos  and  Gaia  DR3  with
predictions  of   our  model  (which   accounts  for  motion   of  the
photo-centre in  both orbits)
and found them to be in reasonable agreement.

\begin{figure}
\includegraphics[width=7cm]{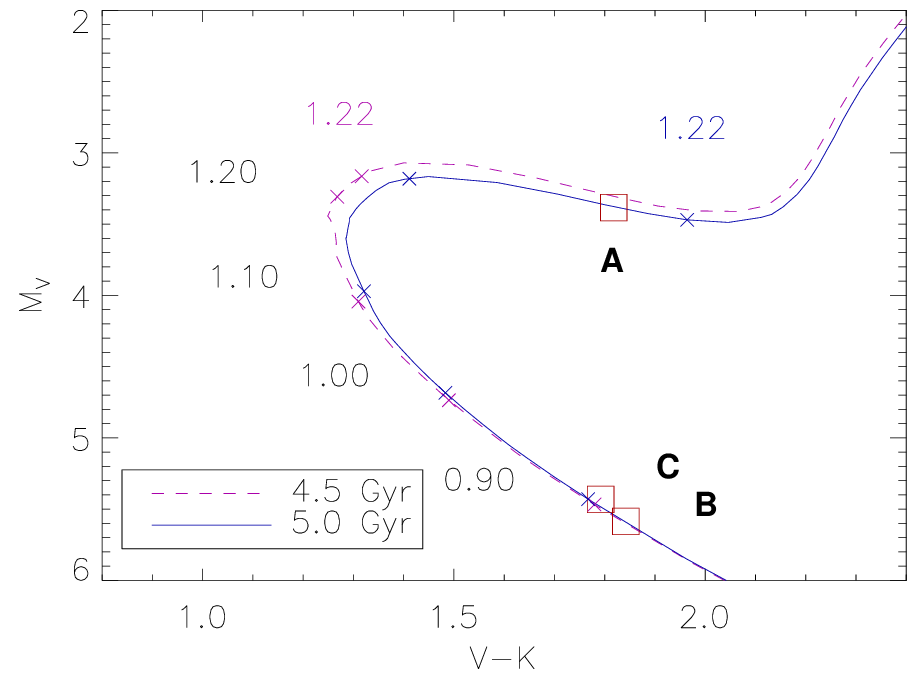}
\caption{Colour-magnitude  diagram  of  the  triple  system  HIP  86286
  (squares)  and PARSEC  isochrones  for 4.5  Gyr and  5  Gyr and  solar
  metallicity, with masses marked.  The mass of A estimated from
    the isochrones is 1.22 \msun.
\label{fig:cmd} 
}
\end{figure}

Figure~\ref{fig:cmd} places the components  of HIP~86286 on the $(M_V,
V-K)$ colour-magnitude diagram. Comparison  with the PARSEC isochrones
\citep{PARSEC}  indicates   an  age   about  5  Gyr.    The  effective
temperatures  and colours  of all  three  stars are  similar, but  the
component A  is substantially brighter than  C and B.  Star  A evolves
rapidly:  a minor  variation of  its mass  or age  results in  a large
change  of the  temperature and  colour; the  two selected  isochrones
bracket the  actual $V-K$  colour at 1.22  \msun ~mass,  slightly less
than 1.28 \msun ~derived by the approximate Montalto formula.

The similarity  of the inclinations and  nodes of the inner  and outer
orbits  is  notable. The  mutual  inclination  between the  orbits  is
$9^\circ$, their  eccentricities are  moderate, and  the ratio  of the
periods  is  only  8.2.   This  triple system  resembles  a  class  of
hierarchies  with relatively  well-aligned and  quasi-circular orbits.
For example,  HIP 85209 (WDS  J17247+3802) is a quadruple  system with
estimated outer period of $\sim$500  yr, intermediate period of 38 yr,
and the  inner (also resolved)  subsystem with  a 1.23 yr  period; its
components have slightly sub-solar masses, and the two inner orbits are
oriented nearly edge-on \citep{Tok2019}.

%-------------------------------------
\section{HIP 117258, a hidden triple system}
\label{sec:117258}

\begin{figure}
\includegraphics[width=7cm]{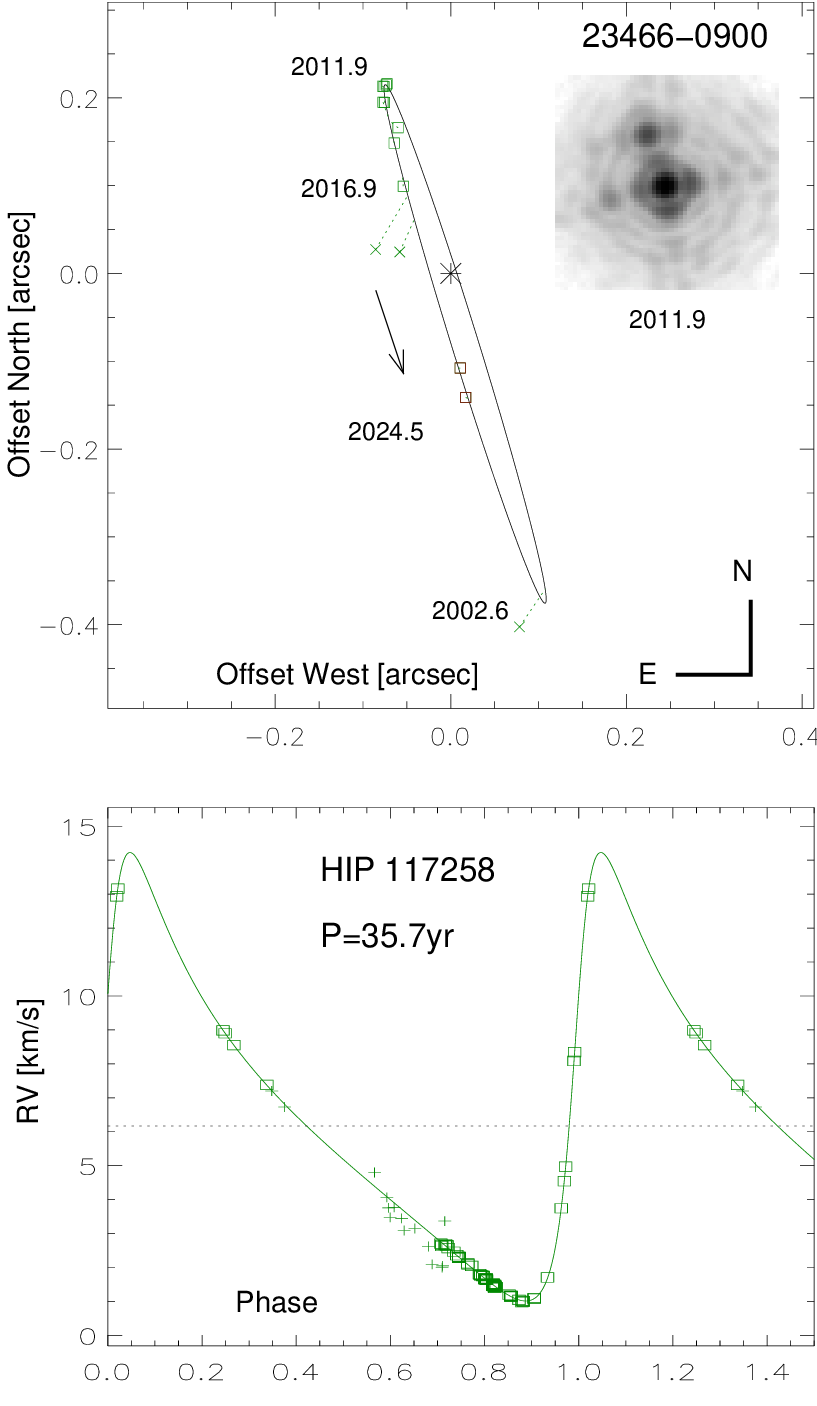}
\caption{Orbit of HIP 117258 in the plane  of the sky (top) and the RV
  curve (bottom; CORAVEL data  are plotted as pluses).  The insert
  shows image of this pair in the $K$ band from \citep{Tok2012}.
\label{fig:orb2} 
}
\end{figure}

\begin{table}
\center
\caption{Orbit and Astrometry of HIP 117258}
\label{tab:orb2}
\medskip
\begin{tabular}{l   c  } 
\hline
Element & Value \\
\hline
$P$ (yr) & 35.71$\pm$0.38  \\
$T$ (JY) & 2008.980$\pm$0.027  \\
$e$      & 0.595$\pm$0.004  \\ 
$a$ ($''$)& 0.3701  \\
$\Omega_{\rm A}$ (deg) & 17.67$\pm$0.20  \\
$\omega_{\rm A}$ (deg) & 291.65$\pm$0.09  \\
$i$ (deg) & 87.55$\pm$0.10  \\
$K_1$ (\kms)  & 6.608 \\
$V_0$ (\kms) & 6.167$\pm$0.030 \\
RA, Decl. (J2000) & 23:46:34.1097, $-$08:59:48.962 \\
$\mu^*_\alpha$, $\mu_\delta$ (\masyr) &  183.66,  $-$68.08 \\
Parallax (mas)    & 26.922$\pm$0.085 \\
\hline
\end{tabular}
\end{table}

This bright ($V=7.228$ mag) solar-type  system, HIP 117258 (HD 223084,
WDS  J23466$-$0900,  LTT  9700,  F9V),  was  identified  as  Hipparcos
accelerating  binary by  \citet{MK05}.   It was  resolved  in 2002  by
\citet{Lloyd2002} with adaptive optics at  the 3-m Lick telescope and,
independently,  in  2011  and  2012 using  the  NICI  adaptive  optics
instrument at Gemini-S \citep{Tok2012,Tok2013}.  The orbital motion of
this  pair was  followed during  2013--2024 by  the speckle  camera at
SOAR.  The  star was also monitored  in RV in search  of exoplanets by
two teams \citep{Fischer2014,Barbato2023}.  A substantial RV variation
indicated a stellar-mass companion, rather  than a planet, and Barbato
et al.  determined a spectroscopic  orbit with  a period of  36.41 yr;
their  RV data  span from  1985  to 2021  and cover  one full  orbital
cycle. They do not mention direct resolutions.

Taken  together,  the  position   measurements  and  the  precise  RVs
constrain  the   orbit  of  this   pair,  seen  edge-on,   quite  well
(Fig.~\ref{fig:orb2}).  The relative photometry of the resolved binary
($\Delta K = 1.86 \pm 0.1$ mag, $\Delta H = 2.19 \pm 0.1$ mag, $\Delta
I =  3.23 \pm 0.3$  mag) indicates that  the secondary companion  B is
fainter and  redder than the main  star A. The IR  photometry suggests
that it should have a mass around 0.65 \msun, assuming that star A has
a mass of 1.1  \msun. However, the predicted $\Delta I  = 2.95$ mag is
smaller than measured.  Furthermore,  the RV amplitude and inclination
imply a  companion of  0.95 \msun;  the large  companion mass  is also
confirmed  by the  measured  acceleration.  If  the  companion were  a
main-sequence  star  with  such  mass,   it  would  be  much  brighter
(estimated  $\Delta  K  =  0.52$  mag and  $\Delta  I  =  0.72$  mag),
contradicting the observations; on the other hand, a white dwarf would
be too faint for imaging detection.

The tension is  resolved by assuming that companion B  is a close pair
of low-mass M-type dwarfs.  The  fitted triple-star model  uses
relative  positions,  RVs,  Hipparcos and  Gaia  astrometry,  combined
photometry, and differential photometry in the $I$ and $K$ bands.  The
orbital elements  in Table~\ref{tab:orb2} result from  this model, and
their errors  are estimated by  MCMC.  Zero extinction is  adopted and
verified.  Parameters of the stars are listed in Table~\ref{tab:par2}.
The modeled magnitude differences between B  and A in the $I$, $H$, and
$K$ bands match the actual differential photometry within errors.  The
data place no strong constraints on the individual masses of Ba and Bb
(only on their sum), and a pair  of equal 0.47 \msun stars would match
observations as well.  

In our  solution this is  constrained by assuming that  the components
are  red  dwarfs,  and  their  masses are  computed  from  absolute  
magnitudes $M_{Ks}$  using the 5th order polynomial  from Table~6 of \citet{Mann2019} for solar
metallicity.
%
%\begin{eqnarray}
%\log M & = -0.642 + M_{Ks} (-0.208 + M_{Ks}(-8.43\,10^{-4}  \nonumber \\ 
%  & + M_{Ks} (7.87\,10^{-3} +  M_{Ks}(1.42\,10^{-4}   \nonumber \\
%  & - 2.13\,10^{-4} M_{Ks} ))))).  
%\label{eq:mann}
%\end{eqnarray}
The mass of  the primary is computed with the  same Montalto's formula
as for all three stars in the first example.
Our dynamical parallax, 26.922 mas, agrees with
the Gaia DR3 parallax of 27.029$\pm$0.081 mas. Naturally, the mass sum
computed from  the orbit with  the Gaia  parallax is also  larger than
1.75 \msun ~expected if it  were a simple binary; conversely, enforcing
such  a small  mass sum  implies  a dynamical  parallax of  28.4\,mas
incompatible with Gaia.

The positions  and residuals are listed  in Table~\ref{tab:pos2}.  The
inaccurate position from  \citet{Lloyd2002} is not used,  but with the
angle changed  by 180$^\circ$,  it fits  the orbit  approximately.  We
also ignore the  tentative measurements at SOAR in  2017.7 and 2018.5,
obtained without  reference stars; the  pair was not resolved  at SOAR
from 2018.7 till 2022.4 and opened up in 2023, after passing through a
conjunction.  We  do not  reproduce here the  published RVs  and their
errors, available in the cited papers.  A slight zero-point correction
was determined for the CORAVEL RVs from \citet{Barbato2023}, and 2.362
\kms was  added to the  RVs from \citet{Fischer2014}, so  our systemic
velocity refers to the CORALIE system.

The  centre-of-mass  PM of  the  system  determined  by our  model  is
$(183.66, -68.08)$ \masyr.  It differs  substantially from the mean PM
derived by \citet{Brandt2021}, $(181.57, -74.67)$ \masyr, because the
orbital period is longer than  24.75 yrs elapsed between Hipparcos and
Gaia DR3  epochs (the companion was  at opposite ends of  its orbit in
those epochs). Our model predicts the photo-centre motion of $(186.95,
-54.28)$ \masyr at the Gaia DR3  epoch and agrees with the measured PM
of $(187.78, -54.37)$ \masyr reasonably well.

\begin{table}
\center
\caption{Stellar Parameters and Photometry of HIP 117258}
\label{tab:par2}
\medskip
\begin{tabular}{l   c  c c c c} 
\hline
Parameter & A  & Ba & Bb &  Ba+Bb & A+B \\
\hline
%Parallax (mas)    & \multicolumn{5}{c}{26.922} \\
Mass (\msun)      & 1.10 &  0.53 & 0.42 & 0.95 & 2.05 \\
Radius ($R_\odot$) & 1.05 &  0.53 & 0.42 & \ldots & \ldots \\
$T_{\rm eff}$ (K)   & 6173 &  3723 & 3497 & \ldots & \ldots \\
$L$ ($L_\odot$)    & 1.44 &  0.049 & 0.024 &  \ldots &   \ldots \\
$\log g$ (cm s$^{-2}$) & 4.440 & 4.708 & 4.809 & \ldots &   \ldots \\
Sp. Type          & F8V & M0.9V & M2.5V & \ldots &   \ldots \\
$B$ (mag)         & 7.764 & 13.760 & 14.921 & 13.440 & 7.76 (7.78) \\
$V$ (mag)         & 7.238 & 12.293 & 13.399 & 11.958 & 7.24 (7.23) \\
$I_C$ (mag)       & 6.636 & 10.304 & 11.088 & 9.874  & 6.636 (\ldots) \\
$J$ (mag)         & 6.224 & 9.149  & 9.795  & 8.672  & 6.12 (6.14)\\
$H$ (mag)         & 6.000 & 8.534  & 9.207  & 8.067  & 5.85 (5.86)\\
$K_s$ (mag)       & 5.939 & 8.313  & 8.962  & 7.837  & 5.77 (5.75)\\
$G$ (mag)         & 7.096 & 11.491 & 12.376 & 11.093 & 7.10 (7.07)\\
$G_{\rm BP}$ (mag)  & 7.408 & 12.555 & 13.669 & 12.222 & 7.40 (7.39)\\
$G_{\rm RP}$ (mag) &  6.711 & 10.492 & 11.275 & 10.062 & 6.66 (6.64)\\
\hline
\end{tabular}
\end{table}

\begin{table}
\center
\caption{Positions and Residuals of HIP 117258}
\label{tab:pos2}
\medskip
\begin{tabular}{l   c cc ccc} 
\hline
Date     & $\theta_{2000}$ & (O-C)$_\theta$ & $\sigma_\theta$ &  $\rho$ & (O-C)$_\rho$ &  $\sigma_\rho$ \\ 
(JY)    & (deg) & (deg)   & (deg) & (mas) & (mas) & (mas) \\
\hline
2002.5877  &  191.0 &$-$5.7 & 999.9   &    410.0 & 34.8  & 999.9 \\    
2011.8483  &   18.9 &$-$0.0 &   2.0   &    227.0 &  1.1  &  5.0  \\ 
2011.8483  &   18.6 &$-$0.3 &   2.0   &    228.0 &  2.1  &  5.0  \\ 
2012.8285  &   19.7 &$-$0.1 &   2.0   &    226.0 &  0.9  &  5.0  \\ 
2012.8285  &   20.0 &  0.2 &   2.0   &    227.0 &  1.9  &  5.0  \\ 
2013.7347  &   21.2 &  0.5 &   1.0   &    208.8 &$-$4.0  &  5.0  \\ 
2014.7613  &   21.6 &$-$0.3 &   1.0   &    209.7 & 19.5  & 50.0  \\ 
2014.8543  &   19.9 &$-$2.1 &   5.0   &    176.4 &-11.5  & 50.0  \\ 
2015.7361  &   23.5 &  0.1 &   1.0   &    161.9 &$-$1.6  &  5.0  \\ 
2016.9568  &   28.6 &  2.4 &   5.0   &    113.2 &-12.2  & 50.0   \\
2017.6802  &   72.3 & 43.4 & 999.9   &     89.9 &-11.6  & 999.9  \\ 
2018.4856  &   67.2 & 33.1 & 999.9   &     63.1 &-11.3  & 999.9  \\ 
2023.4180  &  185.6 &  0.9 &   1.0   &    108.2 &$-$0.2  &  5.0   \\
2024.4666  &  186.9 &$-$1.1 &   1.0   &    142.3 &$-$1.5  &  5.0  \\ 
\hline
\end{tabular}
\end{table}

The triple-star model implies a  magnitude difference of 5 mag between
A and Ba in the $V$ band, so  the 1\% contribution of light from Ba to
the combined spectrum is not  detectable without additional effort. If
B is a close pair of M dwarfs with  a period of a few days, the RVs of
Ba and  Bb vary with large  amplitudes.  Blending of their  weak lines
with the  lines of A  could cause a  subtle variation of  the measured
RVs.   Yet, the  residuals  of  RVs measured  with  CORALIE (the  most
precise data set)  are in agreement with the declared  errors.  The RV
perturbation would  be less if Ba  were a fast rotator  with broad and
shallow lines or if  Ba and Bb were equal and  moved in opposite sense,
canceling potential blending of their lines with A.

A similar solar-type triple system with a dim yet massive companion on
a  26 yr  orbit is  $\kappa$~For or  HIP 11072  \citep{Fekel2018}. The
reality of  the close inner pair  is confirmed by the  X-ray and radio
detections, attributed to active  chromospheres of the low-mass dwarfs
with fast  rotation.  Weak lines  of the secondaries  were tentatively
detected  in the  combined spectrum  after subtracting  the main  star
\citep{KapFor}, suggesting an orbital period of 3.7 days. In contrast,
no X-rays  from HIP 117258 were  reported, probably because it  is 1.6
times further away than HIP 11072.

To search for weak lines of  the inner pair, a high-resolution optical
spectrum of HIP 117258 was taken using the  CHIRON spectrograph at the 1.5-m
telescope  in   Chile  \citep{CHIRON}   on  JD   2,460,543.7706.   Its
cross-correlation with the solar template  reveals only one strong dip
with a heliocentric RV of 6.115  \kms and a width corresponding to the
projected  rotation velocity  of  5.1 \kms.   No  secondary dips  with
amplitude  above 1\%  are found.   The lithium  6707\AA ~line  with an
equivalent width of 50$\pm$5 m\AA ~is present.  The projected rotation
velocity  corresponds  to  a  rotation   period  of  10  days  if  the
inclination  is  close  to  90$^\circ$,  or  shorter  if  the  axis  is
inclined. The RV agrees  with the orbit (residual +0.16 \kms).

A close pair  in the secondary component of HIP  117258 could manifest
itself by flux modulation due  to eclipses or starspots.  However, the
light curves recorded by the TESS satellite \citep{TESS} do not reveal
any eclipses. Instead,  an irregular modulation with a period  of 5 or
10  days and  an amplitude  of $\sim$0.3\%  is seen.   Considering the
small contribution of Ba and Bb  to the total flux, this modulation is
likely caused  by starspots  on the  main star.  No strong  flares were
noted.

Summarizing,  the case  for  HIP  117258 hosting  a  triple system  is
strong, and there are no other  viable explanations for the large mass
of the dim  secondary. The evidence, however, is  indirect. Our
efforts to find additional clues from spectroscopy or space photometry
have not produced any tangible results.
 
%-------------------------------------
\section{Summary}
\label{sec:sum}

We  presented here  the method  where diverse  data available  on each
stellar  system are  fitted by  a model  comprising orbital  elements,
physical  parameters  of  the  stars,   and  distance 
(dynamical  parallax).  The  assumption   that  stars  obey  empirical
relations between mass, effective temperature, and radius is built
into the model and checked by examination of the fitted parameters.

The method is applied here  to two nearby triple systems, establishing
new  facts on  their structure  and  parameters.  The  first one,  HIP
86286,  is  unusual because  its  two  faint resolved  companions  are
located at comparable separations. We determine the orbital periods of
35 and 287 years and show  that the orbits are approximately coplanar.
This system  therefore has  a planetary-type architecture.   The other
system,  HIP 117258,  has been  known until  now as  a  simple binary.   We
determine  its  accurate  36-yr  orbit, revealing  that  the  secondary
companion is  over-massive. The  secondary should be  a close  pair of
M-type dwarfs with yet unknown period.  So, HIP 117258 joins the group
of triple solar-type  stars with dim yet  massive secondary companions
which are themselves close pairs, similar to $\kappa$~For (HIP 11072).

The two  triple  systems  studied  here are  problematic  for  Gaia,  for
different reasons (complicated source  structure and insufficient time
coverage), but  the combination of  Gaia with ground-based data  has a
great power.

Our  findings  contribute  to  the database  on  hierarchical  systems
\citep{MSC},  albeit  by a  small  increment,  and advance  the  still
incomplete  knowledge  of  multiple-star   populations  in  the  solar
neighborhood.

%-------------------------------------
%\section{}
%\label{sec:}
%-------------------------------------
%\section{}
%\label{sec:}
%-------------------------------------
%\section{}
%\label{sec:}
%-------------------------------------
%\section{}
%\label{sec:}

%--
\section*{Acknowledgments}

 We thank the Referee for useful comments that improved the paper.
This work  used the  SIMBAD service operated  by Centre  des Donn\'ees
Stellaires  (Strasbourg, France),  bibliographic  references from  the
Astrophysics Data  System maintained  by SAO/NASA, and  the Washington
Double  Star  Catalog  maintained  at USNO.  We  thank  R.~Matson  for
retrieving  the  historic  micrometer measurements.   Modern  position
measurements  used here  were obtained  at the  Southern Astrophysical
Research  telescope.  This work has  made use of data  from the
European       Space       Agency       (ESA)       mission       Gaia
(https://www.cosmos.esa.int/gaia),   processed   by  the   Gaia   Data
Processing        and         Analysis        Consortium        (DPAC,
https://www.cosmos.esa.int/web/gaia/dpac/consortium).
\section*{Data Availability}

Only published data  were used in  this research. 

%------------------------------------------------------------------------

%------------------------------------------------------------------------
%\appendix

%------------------------------------------------------------------------
%------------------------------------------------------------------------

\bsp

\label{lastpage}

\end{document}